# Realization of second-order photonic square-root topological insulators

Wenchao Yan[1], Daohong Song[1,2*], Shiqi Xia[1], Junfang Xie[1], Liqin Tang[1], Jingjun Xu[1,2*] and Zhigang Chen[1,2,3*]

[1]*Key Laboratory of Weak-Light Nonlinear Photonics, Ministry of Education, TEDA Institute of Applied Physics and School of Physics, Nankai University, Tianjin 300457, China*

[2]*Collaborative Innovation Center of Extreme Optics, Shanxi University, Taiyuan, Shanxi 030006, China*

[3]*Department of Physics and Astronomy, San Francisco State University, San Francisco, California 94132, USA*

*Corresponding authors: songdaohong@nankai.edu.cn, jjxu@nankai.edu.cn, zgchen@nankai.edu.cn*

**Abstract:** Square-root higher-order topological insulators (HOTIs) are recently discovered new topological phases, with intriguing topological properties inherited from a parent lattice Hamiltonian. Different from conventional HOTIs, the square-root HOTIs typically manifest two paired non-zero energy corner states. In this work, we experimentally demonstrate the second-order square-root HOTIs in photonics for the first time to our knowledge, thereby unveiling such distinct corner states. The specific platform is a laser-written decorated honeycomb lattice (HCL), for which the squared Hamiltonian represents a direct sum of the underlying HCL and breathing Kagome lattice. The localized corner states residing in different bandgaps are observed with characteristic phase structures, in sharp contrast to discrete diffraction in a topologically trivial structure. Our work illustrates a scheme to study fundamental topological phenomena in systems with coexistence of spin-1/2 and spin-1 Dirac-Weyl fermions, and may bring about new possibilities in topology-driven photonic devices.

Topological insulators have attracted increasing attention because of their intriguing fundamental physics and potential applications[1], particularly in the realm of photonics[2]. In fact, photonic topological insulators and associated development of laser technologies[3] have become one of the leading frontiers in optics and photonics, going froward to the next decade[4]. A key feature of the topological insulators is the existence of gapless edge states, which exhibit robust unidirectional transport without backscattering even in the presence of defects or disorders[5-7]. According to the normal bulk-boundary correspondence, a conventional *d*-dimensional topological insulator supports (*d*-1)-dimensional boundary states. However, the newly discovered higher-order topological insulators (HOTIs) can support not only (*d*-1) but also ($d$-$n$)-dimensional boundary states (with $n \geq 2$; a typical example is the 0D corner states in a 2D system)[8-12]. Thus far, HOTIs have been experimentally demonstrated in condensed matter systems[13] as well as a variety of other synthetic platforms ranging from electric circuits[14-16] to acoustics[17-20] and photonics[21-27]. In addition to the fundamental interest, HOTIs are highly tested and touted for novel applications in robust photonic crystal nanocavities[28] and low-threshold topological corner state lasing[29,30]. Indeed, research activities on HOTIs have blossomed[31,32], from linear to nonlinear regimes[33-37], from real to synthetic dimensions[38], and from Hermitian to non-Hermitian systems[39].

In 2017, a novel class of topological insulators, namely, the square-root topological insulators, was proposed, where a theoretical recipe was used to construct a lattice with topological states inherited from the nontrivial square root of its parent lattice Hamiltonian, akin to the passage taken from the Klein-Gordon equation to the Dirac equation[40]. Recently, several approaches for generating and generalizing the square-root Hamiltonian of a given topological insulator have been proposed[41-45]. The square-root topological insulators can support multiple topological edge states as the number of bandgaps doubled compared with that supported by the original parent Hamiltonian. Such square root topological insulators have been experimentally realized in 1D photonic systems[46,47]. Moreover, the topological skin effect in non-Hermitian 1D photonic square root topological insulators was also theoretically studied[48]. Interestingly, the notion of the square root operation can also be extended to HOTIs, where the emergence of paired in-gap corner states in square-root HOTIs was found to be mediated by higher-order topology of the parent Hamiltonian[48-50]. So far, however, the square root HOTIs have been experimentally demonstrated in electric circuits and acoustic structures[50-52], but not in photonics which offers a new scheme for robust light guiding and trapping. Moreover, it is worth noting that the concept of square-root topology has been extended to $2^n$-root topological insulators, topological semimetals, and other HOTI creatures[53,54].

In this work, we experimentally demonstrate the first photonic square-root HOTI. The photonic platform relies on judiciously designed lattice structures – the so-called decorated honeycomb lattices (HCLs) illustrated in Fig 1(a), which simultaneously possess a Lieb-like pseudospin-1 and several graphene-like pseudospin-1/2 Dirac-like cones. The higher-order topological properties of such a system are inherited from one of the parent Hamiltonian which contains the breathing Kagome lattice. We establish the decorated HCLs by cw-laser writing waveguides in the bulk of a nonlinear crystal, which offers exquisite control over initial conditions, and thereby directly identify two types of topological corner states dwelling in different bandgaps by examining their phase structures. Wave dynamics in topologically trivial and nontrivial decorated HCLs are directly compared in both experiments and numerical simulations.

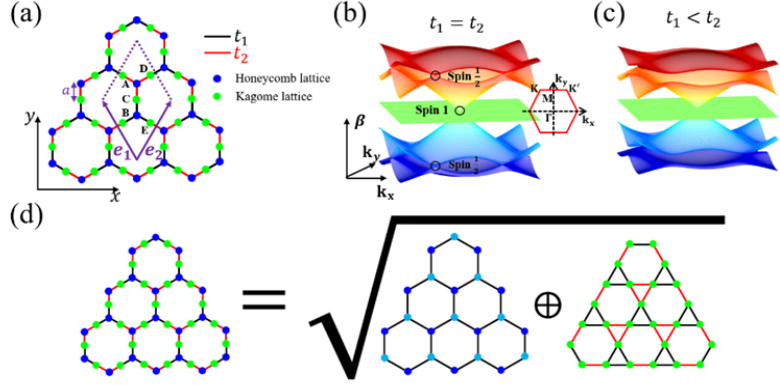

**Fig. 1.** (a) Schematic of a decorated HCL structure consisting of five sites (A, B, C, D, E) per unit cell shown in a purple rhombus, where $e_1$ and $e_2$ illustrate the lattice vectors. (b) - (c) Calculated band structure from the tight-binding model for (b) $t_1 = t_2 = 1$, and (c) $t_1 = 0.2$, $t_2 = 0.8$. (d) Illustration of the relation between the original Hamiltonian of the decorated HCL and the square root of its parent lattices.

The decorated HCL[49,52] (a somewhat mixed-use term with edge-centered HCL[55], super HCL[56,57] and hybrid HCL[51]) consists of five lattice sites (A, B, C, D, E) per unit cell as marked by a dark-dashed square in Fig. 1(a), which can be considered as a combination of a honeycomb lattice (A, B) and a Kagome lattice (C, D, E). A light beam propagating in such a 2D photonic decorated HCL is governed by the Schrödinger-type paraxial wave equation[58,59]:

$$i\frac{\partial \Psi(x,y,z)}{\partial z} = -\frac{1}{2k_0}\nabla^2 \partial \Psi(x,y,z) - \frac{k_0 \Delta n(x,y)}{n_0}\Psi(x,y,z) \equiv H_0 \Psi(x,y,z) \qquad (1)$$

Here $\Psi$ is the electric field envelope of the probe beam, $\nabla^2 = \partial_x^2 + \partial_y^2$ is the transverse Laplacian operator, $z$ is the longitudinal propagation distance in the photonic lattice, $k_0$ is the wavenumber in the medium, $n_0$ is the background refractive index, and $\Delta n$ is the refractive index change that defines the photonic decorated HCL. $H_0$ in Eq. (1) is the continuum Hamiltonian for wave propagation in the photonic lattice. In 2020, T. Mizoguchi *et al.* theoretically introduced how to demonstrate the novel square root topological insulators in the decorated HCL[49]. If we only consider the nearest-neighbor coupling of the waveguides and define two different hopping parameters $t_1$ and $t_2$, we can get the corresponding tight-binding Hamiltonian in $k$-space[49]

$$H_T = \begin{bmatrix} O_{2,2} & \Phi_T^\dagger \\ \Phi_T & O_{3,3} \end{bmatrix} \qquad (2)$$

Here $O_{m,n}$ represents the $m \times n$ zero matrix, the $\Phi_T$ is the 3×2 matrix

$$\Phi_T = \begin{bmatrix} t_2 d_1 & t_1 d_1^* \\ t_2 d_2 & t_1 d_2^* \\ t_2 d_3 & t_1 d_3^* \end{bmatrix} \quad (3)$$

where $d_1 = \exp(ik_y a)$, $d_2 = \exp[-i(\frac{\sqrt{3}}{2}k_x a + \frac{1}{2}k_y a)]$, $d_3 = \exp[i(\frac{\sqrt{3}}{2}k_x a - \frac{1}{2}k_y a)]$, and $a$ is the lattice constant. Diagonalizing $H_T$ yields the standard band structure $\beta(k)$ of decorated HCL when $t_1 = t_2$ as shown in Fig. 1(b). In this case, there exist two kinds of Dirac-like cones: one is a Lieb-like pseudospin-1 Dirac-like cone formed by a flat-band touching with two dispersive bands at the Γ point of the first Brillouin Zone (BZ), and the other is the graphene-like pseudospin-1/2 Dirac cones appearing at K and K' points. However, for the pseudospin-1/2 Dirac cones, there will be opened gaps at K and K' points when the decorated HCL is deformed with $t_1 < t_2$ as shown in Fig. 1(c). In such a system, it is possible to have conventional topological edge states and higher-order topological corner states in the two different gaps. Interestingly, the Hamiltonian $H_T$ is chiral-symmetric, and satisfies the following relationship: $\gamma H_T \gamma^\dagger = -H_T$, where

$$\gamma = \begin{pmatrix} I_{2,2} & O_{2,3} \\ O_{3,2} & -I_{3,3} \end{pmatrix} \quad (4)$$

Here $I_{m,m}$ represents the $m \times m$ identity matrix. This indicates the existence of the parent Hamiltonian whose square root corresponds to $H_T$. So we can take the square of $H_T$ as follows:

$$[H_T]^2 = \begin{bmatrix} h_T^{(H)} & O_{2,3} \\ O_{3,2} & h_T^{(K)} \end{bmatrix} \quad (5)$$

Here $h_T^{(H)} = \Phi_T^\dagger \Phi_T = \begin{bmatrix} 3t_2^2 & t_1 t_2 d_4^* \\ t_1 t_2 d_4 & 3t_1^2 \end{bmatrix}$, $h_T^{(K)} = \Phi_T \Phi_T^\dagger = \begin{bmatrix} t_1^2 + t_2^2 & t_2^2 d_5 + t_1^2 d_5^* & t_2^2 d_6 + t_1^2 d_6^* \\ t_2^2 d_5^* + t_1^2 d_5 & t_1^2 + t_2^2 & t_2^2 d_7 + t_1^2 d_7^* \\ t_2^2 d_6^* + t_1^2 d_6 & t_2^2 d_7^* + t_1^2 d_7 & t_1^2 + t_2^2 \end{bmatrix}$.

$d_4 = \exp(2ik_y a) + \exp[-i(\sqrt{3}k_x a + k_y a)] + \exp[i(\sqrt{3}k_x a - k_y a)]$, $d_5 = \exp[i(\frac{\sqrt{3}}{2}k_x a + \frac{3}{2}k_y a)]$, $d_6 = \exp[-i(\frac{\sqrt{3}}{2}k_x a - \frac{3}{2}k_y a)]$, $d_7 = \exp(-i\sqrt{3}k_x a)$. For $h_T^{(H)}$, one can see that the coupling coefficients between different sublattice sites in the off-diagonal terms are the same, but the diagonal terms are different which are determined by $t_1$ and $t_2$. However, for $h_T^{(K)}$, one can see that the diagonal terms are the same while the off-diagonal terms are different if $t_1$ is not equal to $t_2$. As such, the $h_T^{(H)}$ represents the Hamiltonian of honeycomb lattice with different sublattice on-site potentials, and $h_T^{(K)}$ represents the Hamiltonian of the breathing Kagome lattice as shown in Fig. 1(d), which has been illustrated as a typical example of HOTIs[20,21,26] although with much debate[60,61].

In this work, we design and establish finite photonic decorated HCLs with 81 lattice sites under open boundary conditions as depicted in Fig. 2(a) and 2(b). The choice of the terminations is that all sites in the sublattice of decorated HCL that become $h_T^{(K)}$ upon squaring must have the same number of connections ($t_1$ and $t_2$), such that the diagonal term in $h_T^{(K)}$ is uniform. Here, $S_1$ and $S_2$ mark two different lattice spacings between nearest neighbor lattice sites, which determine the nearest neighbor waveguide coupling $t_1$ and $t_2$. Experimentally, by controlling the waveguide spacing $S_1$ and $S_2$, the decorated HCL can be nontrivial (or trivial) if $S_1 > S_2$ (or $S_1 < S_2$). In the tight-binding model, we can calculate the eigenvalues of the decorated HCL as a function of the dimerization parameter $c$ ($c = t_1 - t_2$) as shown in Fig. 2(c), the topological characteristics of a decorated HCL are discussed from the emergence of corner and edge states in nontrivial lattices where $c$ is smaller than zero (i.e. $t_1 < t_2$). In order to study the HOTIs in the nontrivial decorated HCL, we calculated the energy spectrum, the corner and edge states for $t_1 = 0.2$, $t_2 = 0.8$ as shown in Figs. 2(d)-2(h). There exist two types of three-fold degenerate corner states as illustrated in Fig. 2(e) and 2(f), which occupy all three corner sites, and the nearest neighbor sites of the corner states (Fig. 2(e)) in the top gap are out of phase, as the phase difference between adjacent neighboring sites are $\pi$, thus named as "out-of-phase corner states". For the corner states (Fig. 2(f)) in the bottom gap, however, the three neighboring corner sites all have the same phase, so they are named as "in-phase corner states". These two different in-gap corner states have distinct phase relation among the nearest neighbor sites, which can be directly observed in experiment (as will be shown below). Intriguingly, these two kinds of corner states reside on both the Kagome and the honeycomb sublattices as shown in Fig. 2(e) and (f), although it is expected that the higher order topology in such a system should be inherited from the breathing Kagome lattice[49].

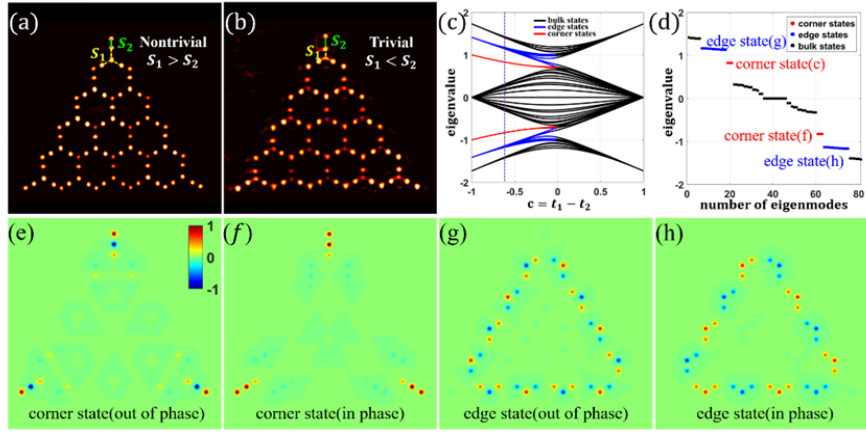

**Fig.2.** (a)(b) Experimentally established (a) nontrivial and (b) trivial photonic decorated HCLs by use of the cw-laser-writing technique. $S_1$ and $S_2$ mark two different lattice spacings of nearest neighbor lattice sites which determine the waveguide coupling $t_1$ and $t_2$, respectively. (c) Calculated eigenvalues of the decorated HCL as a function of the dimerization parameter $c$, $c = t_1 - t_2$, the corner and edge states are highlighted with red and blue colors in the topologically nontrivial regime. (d) The energy spectrum for a nontrivial decorated HCL with c=-0.6 ($t_1 = 0.2$, $t_2 = 0.8$). The red, blue, and black circles represent the corner, edge and bulk states, respectively. (e)(f) Mode profiles of (e) out-of-phase corner states in the top gap and (f) in-phase corner states in the bottom gap. (g)(h) Mode profiles of the edge states in the top and bottom gaps, respectively.

Next, we present our experimental results of the predicted corner states. The experimental setup is similar to our recent work on the observation of flatband line states in finite photonic decorated HCLs[57], which are established by site-to-site waveguide "writing" in a non-instantaneous nonlinear strontium barium niobate (SBN:61) crystal with a continuous-wave laser beam. We experimentally established two kinds of deformed decorated HCLs by changing the waveguide spacing across the entire lattice, one has topologically nontrivial termination (Fig. 2a) and the other one has trivial termination (Fig. 2b). For our experiment, the two parameters for waveguide spacing in the nontrivial (trivial) decorated HCL are $S_1$=42μm, and $S_2$=30μm ($S_1$=30μm, and $S_2$=42μm). For these experimental lattice parameters, we estimated numerically that the corresponding coupling coefficient ratio $t_1/t_2$ is about 0.19, which is lower than that used in calculation for Fig. 2, but a slight difference in coupling ratio in the topological nontrivial regime will not affect the corner mode structure and the physics about the square-root HOTI presented here (see Supporting Information). In our experiment, the different corner states are excited by a dipole-like probe beam but with different phase structures to match the modes of the corner states. Specifically, for excitation of the out-of-phase corner states, the input dipole-like beam has out of phase structure, instead, the in-phase corner states are excited by the in-phase dipole-like beam. To observe the out-of-phase corner state in the top gap as illustrated in Fig. 2(e), we use a spatial light modulator (SLM) to obtain an out-of-phase dipole-like beam, then employ it to excite the two sites at the left bottom corner as illustrated in Fig. 3(a). The probe beam is mainly localized at the initially excited corner sites but with evident coupling to the third site from the corner (i.e., the nearest neighbor site) in the nontrivial decorated HCL after 2 cm of propagation, as illustrated in Fig. 3(b2). Furthermore, in order to measure the phase structure, we use an inclined broad beam (quasi-plane wave) to interfere with the output beam to get the interferogram. The interference fringes between neighboring sites give rise to information about their phase difference, as commonly used in optics experiments: if the fringes are lined up, then the phase difference between the adjacent sites is 0, i.e., they are in phase; however, if the fringes are staggered or interleaved, the phase difference between the adjacent points is $\pi$, i.e., they are out of phase. In Fig. 3(b3), from the interferogram, it is clear that the adjacent three sites of the output pattern have opposite phase, which agrees with predicted phase structure of the corner state in the top bandgap as shown in Fig. 2(e). For direct comparison, we perform similar experiment in a trivial photonic decorated HCL as illustrated in Fig. 2(b), and the corresponding results are presented in Fig. 3(c2) and Fig. 3(c3). In this case, the probe beam couples not only into the nearest neighbor site, but also into the next nearest neighbor sites and beyond. More importantly, the interferogram shows that the output has in phase structure with the nearest neighbor site at the corner [Fig. 3(c3)]. These results indicate that we can only observe the corner states in the nontrivial decorated HCL, not in the trivial one. In order to further confirm that the corner localization in nontrivial decorated HCL, similar experiments were further conducted in the bulk of nontrivial lattice with other parameters unchanged, and the corresponding results are shown in Fig. 3(d). One can see that the probe beam also couples to the nearest neighbor sites, whose phase structure is different from that of the corner state in Fig. 3(d3) as the light field in the yellow and white sites are in phase. Due to the limited propagation distance in our experiment (2 cm), we also perform numerical simulations with similar parameters for much longer propagation distance (20 cm) so to better appreciate the experimental results as shown in Fig. 3(b4) -Fig. 3(d4). The differences for the three cases under long distance propagation are more obvious: only the probe beam excitation at the corner of the nontrivial decorated HCL remains localized due to the excitation of the corner state [Fig. 3(b4)], while that for other two cases it spreads more into the bulk [Fig. 3(c4) and Fig. 3(d4)]. These results clearly demonstrate the observation of the out-of-phase corner state in the nontrivial photonic decorated HCL.

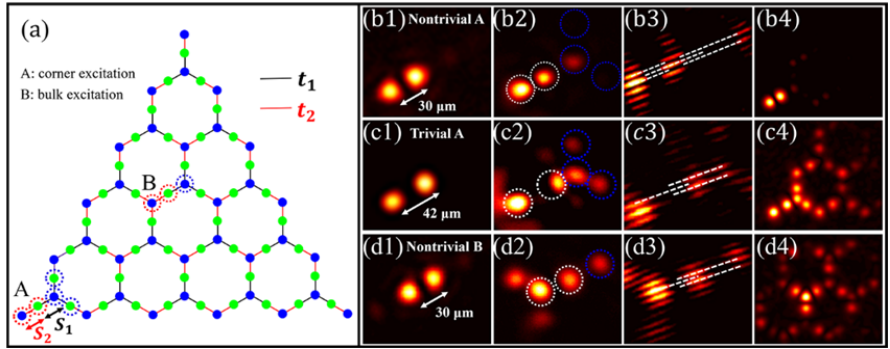

**Fig. 3**. Experimental demonstration of out-of-phase corner states in photonic decorated HCLs. (a) Schematic of the decorated HCL structure, where the red dashed circles mark the excitation positions of the dipole-like probe beam for corner (A) and bulk (B) excitations under trivial ($S_1<S_2$) and nontrivial ($S_1>S_2$) conditions. Right panels show experimental (middle three columns) and simulation (last column) results under different initial excitation conditions with an out-of-phase dipole-like probe beam. (b) and (c) correspond to corner excitation in (b) nontrivial and (c) trivial decorated HCLs, respectively, while (d) corresponds to bulk excitation in the nontrivial lattice, as illustrated in (a). From left to right, shown are input intensity pattern, corresponding output exiting the lattice, zoom-in interferogram of the output pattern, and simulation results of the output intensity distribution after a much longer propagation distance (20 cm).

Similarly, we have experimentally observed the in-phase corner states in the bottom energy bandgap as illustrated in Fig. 2(f). In this case, we sent an in-phase dipole-like beam into the corner of the nontrivial decorated HCL, and the output results of the probe beam are illustrated in Figs. 4(a2-b2). It is clear that the probe beam is localized mainly at the two corner waveguides initially excited in the nontrivial decorated HCL (Fig. 4(a2)), and the interference pattern obtained with an inclined broad beam confirms the in phase structure (Fig. 4(a3)). The output intensity pattern can remain intact for much longer propagation distance (20 cm) from simulation with similar parameters to our experiment as shown in Fig. 4(a4) due to the excitation of the in-phase corner states, while the same in-phase dipole beam can not be localized at the corner sites in the trivial decorated HCL as presented in Figs. 4(b2)-4(b4). It should be noted that the in-phase corner states do not exist in conventional breathing Kagome lattice, where only the out-of-phase corner states exist. Such in-phase corner state is a unique feature of the nontrivial decorated HCL arising from square-root higher order topology.

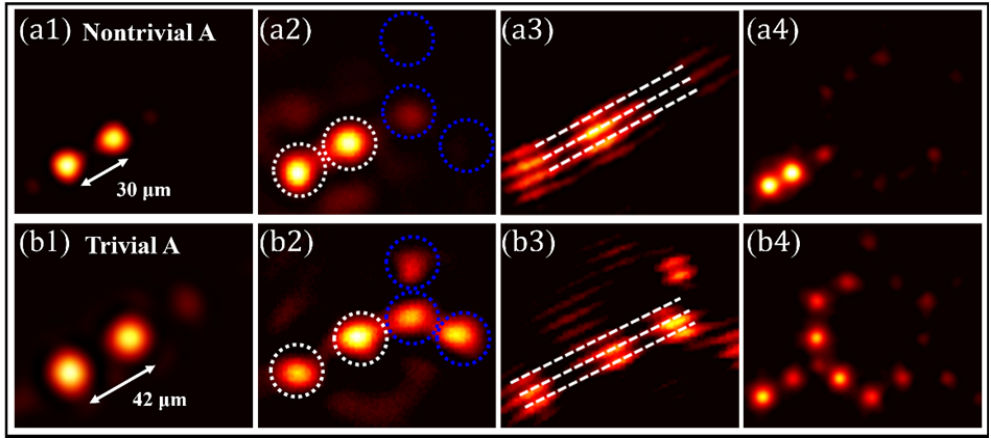

**Fig.4.** Experimental observation of the in-phase corner states in photonic decorated HCLs. (a) and (b) have the same layout as that of (b) and (c) in Fig. 3, except now the input two spots (a1, b1) forming the dipole-like excitation beam are in phase. Experimental results are shown in the left three columns, and the last column is long-distance (20 cm) simulation results, for corner excitation in (a) nontrivial and (b) trivial decorated HCLs. Notice the line-up fringes (in-phase relation) in (a3, b3) as opposed to the interleaved fringes in Fig. 3(b3, c3) (out-of-phase relation).

Before conclusion, we would like to mention the ongoing debate about the breathing Kagome lattices being characterized as nontrivial HOTIs[60,61]. It was argued that there is a fundamental difference between the chiral symmetry topological crystalline insulators with even-fold symmetry (such as the 2D SSH structure[11,12,14,22,23,25,36] and the one that does not (such as the breathing Kagome lattices with $C_3$ symmetry[17,20,21,26,37]). As demonstrated recently, even though the chiral symmetry on its own is insufficient to stabilize the corner modes against strong edge perturbations, additional presence of the rotational symmetry (such as $C_{4v}$ exhibited in 2D SSH lattices) can offer topological protection and further entail the formation of topological corner-localized states. However, it remained elusive whether the previously observed corner states in the breathing Kagome lattices[17,20,21,26,37] can be strictly classified as the HOTI states[60,61]. While we have observed clearly the corner states in the decorated HCL, whose topological features are inherited from the parent Hamiltonian of the breathing Kagome lattice, further theoretical analysis following similar approach of Refs. [60,61] is desired for rigorously affirming the topological phase and understanding the observed corner states in the square-root HOTIs.

In conclusion, we have experimentally demonstrated for the first time second-order square-root HOTIs in a photonic decorated HCL platform. We have observed two kinds of corner states which reside in different energy band gaps with different phase structures. Our work not only provides a new platform to study the higher-order topological physics in optics, but also brings about new possibilities on future studies of other novel HOTIs, for both fundamental understandings and applications. For example, the recently proposed $2^n$-root 2D HOTIs such as the "quartic-root" version of the breathing Kagome lattices (described by the quartic-root HOTIs tight-binding Hamiltonian[54]) can be implemented in judiciously designed photonic lattices, in which multiple distinct corner states with different phase structures can also be excited and probed by using the similar experimental methods present in our work.

Note added in proof: we became aware that a manuscript reporting related results was just submitted to arXiv [62].

**Supporting Information:**
Numerical simulation and discussion of the coupling coefficients.

**Acknowledgments.** This work was supported by National Key R&D Program of China (2017YFA0303800); the Chinese National Science Foundation (11922408, 11674180 and 91750204); the Fundamental Research Funds for the Central Universities(63213041,63211001), PCSIRT (IRT_13R29), 111 Project (No. B07013) in China.

**Disclosures.** The authors declare no conflicts of interest.